\documentclass{article}
\usepackage{spconf,amsmath,graphicx}

\usepackage{enumitem}
\setlist{nosep, leftmargin=14pt}

\usepackage{mwe} 
\usepackage{amssymb}
\usepackage{multirow}

\title{Improving Uncertainty-based Out-of-Distribution detection for medical image segmentation}

\name{Benjamin Lambert$^{1,3}$, Florence Forbes$^{2}$, Senan Doyle$^{1}$, Alan Tucholka$^{1}$, Michel Dojat$^{3}$}

\address{
$^{1}$ Pixyl, Research and Development Laboratory, 38000 Grenoble, France \\
$^{2}$ Univ. Grenoble Alpes, Inria, CNRS, Grenoble INP, LJK, 38000 Grenoble, France \\
$^{3}$ Univ. Grenoble Alpes, Inserm, U1216, Grenoble Institut Neurosciences, GIN, 38000 Grenoble, France}

\begin{document}
%
\maketitle
\begin{abstract}

Deep Learning models are easily disturbed by variations in the input images that were not seen during training, resulting in unpredictable behaviours. Such Out-of-Distribution (OOD) images represent a significant challenge in the context of medical image analysis, where the range of possible abnormalities is extremely wide, including artifacts, unseen pathologies, or different imaging protocols. In this work, we evaluate various uncertainty frameworks to detect OOD inputs in the context of Multiple Sclerosis lesions segmentation. By implementing a comprehensive evaluation scheme including 14 sources of OOD of various nature and strength, we show that methods relying on the predictive uncertainty of binary segmentation models often fails in detecting outlying inputs. On the contrary, learning to segment anatomical labels alongside lesions highly improves the ability to detect OOD inputs. 

\end{abstract}
\section{Introduction}
\label{sec:intro}

\subsection{Motivations}
Out-of-distribution (OOD) images are defined as being far from the training dataset distribution. Deep Learning (DL) models tend to behave inconsistently for this type of inputs, making OOD image detection crucial to avoid hidden model deficiencies \cite{yang2021generalized}. It is especially required in real-world automated pipelines, where input images may not be visually inspected before running the analysis. In the context of computerized medical-images analysis, a large variety of phenomenons in the input images can disturb a DL model and lead to unpredictable behaviours: noise, artifacts, variations in the imaging acquisition protocol and device, or pathological cases that were not included in the training dataset.

\subsection{Related Works}
OOD image detection frameworks can roughly be divided into two categories \cite{berger2021confidence}: methods that build a model specifically dedicated to the OOD task, and methods that rely on the prediction uncertainty of a trained model to detect abnormal inputs. 

Within the first category, the most straightforward manner is to build a classifier directly on input images for the detection of OOD ones. For this, a Convolutional Neural Network (CNN) can be trained in a supervised manner, using an annotated dataset containing various types of real-world OOD \cite{bottani2022automatic}. Alternatively, Unsupervised Anomaly Detection (UAD) proposes to model the appearance of normal images by training an Auto-Encoder network (AE) to reconstruct in-distribution (ID) samples. At test-time, reconstruction is expected to be degraded for OOD samples, allowing for their detection \cite{gong2019memorizing,park2020learning}. Finally, in the context of semantic image segmentation, the Image Resynthesis framework \cite{lis2019detecting,xia2020synthesize} proposes to train a Generative Adversarial Network (GAN) to reconstruct the input image from the predicted segmentation. As the automated segmentation of OOD images is expected to be poor, the resynthesized input image should be distant from the true (unseen) input image.

While the former methods require the training of an auxiliary model (CNN, AE, or GAN) dedicated to the OOD detection task, uncertainty quantification (UQ) approaches propose to detect OOD inputs directly from the outputs of the trained predictive neural network. They rely on the hypothesis that the uncertainty of the deployed model should be higher when confronted to an unusual image, compared to the one of ID images. Based on this principle, OOD detection can be performed by monitoring the output uncertainty of the model. Various frameworks have been proposed to quantify uncertainty, and we present four of the most popular approaches below. 

The baseline approach to estimate the uncertainty of a trained prediction model is to consider its output softmax probabilities. Following the intuition that the higher the probability, the more certain is the model, the \emph{Maximum Softmax Probability} (MSP) can be used as an estimator of uncertainty \cite{hendrycks2016baseline}. Another standard UQ methodology consists in producing a set of diverse and plausible predictions for the same input image. To achieve that, the \emph{MC dropout} approach \cite{gal2016dropout} consists in training a segmentation model with dropout and keep it activated during inference. At test time, each input image is forwarded multiple times through the model, with randomly sampled Dropout masks, yielding to multiple predictions for the same input image. Alternatively, \emph{Deep Ensemble} (DE) \cite{lakshminarayanan2017simple} proposes to train the same neural network multiple times. As the weights are randomly initialized at each training, the models reach different optimums. At test time, each individual model processes the input image independently, which produces a set of predictions. For both the MC dropout model and DE, uncertainty can then be estimated by computing the variance among the predictions. Finally, \emph{Deterministic Uncertainty Methods} (DUM) analyze the intermediate feature maps of a trained model to detect OOD inputs \cite{postels2021practicality}. It is based on the assumption that the hidden activations of the model should be different between an ID sample and an OOD one. In \cite{karimi2020improving}, authors propose to analyze the signature of an input image in the latent space of a segmentation model to detect OOD samples. At test time, an image OOD score is obtained by computing the distance between the signature of the test image, and the signatures of all training images.

As mentioned, OOD detection in medical image segmentation is critical, particularly in a clinical setting. Although uncertainty-based OOD detection has been applied for medical image segmentation \cite{karimi2020improving,mehrtash2020confidence,diao2022unified}, its evaluation is usually limited to extreme OOD cases (\textit{e.g.} different imaging modality, organ of interest or imaging center). While it is necessary to robustly detect such cases, we believe that the detection of more subtle artifacts (\textit{e.g.} image artifacts, truncation or skipped preprocessing) is also crucial in clinical practice. 

\subsection{Contributions}
In this work, we propose to further explore uncertainty-based OOD detection in the context of medical image segmentation. As a use-case, we study OOD detection in input MRIs in the context of Multiple Sclerosis (MS) lesions segmentation. We develop a robust evaluation framework containing 14 synthetic or real abnormalities, representative of the abnormalities encountered in clinical routine. We use this framework to compare four of the most popular UQ-based OOD-detection models. Each model, initially binary, is additionally derived as a multiclass model that also segments anatomical regions of the brain in parallel with the MS-lesion segmentation. Indeed, we hypothesize that binary models trained to solely segment lesions tend to discard information not relevant for the task at hand, which may leave out key information for OOD detection. To verify this intuition, we compare the binary and multiclass versions of each segmentation model with respect to the OOD detection task.
We demonstrate that the retention of all additional information in multiclass models highly improves the OOD performance of uncertainty-based techniques, as compared to binary models. 

\begin{figure*}[!t]
\centering
\includegraphics[width=\textwidth]{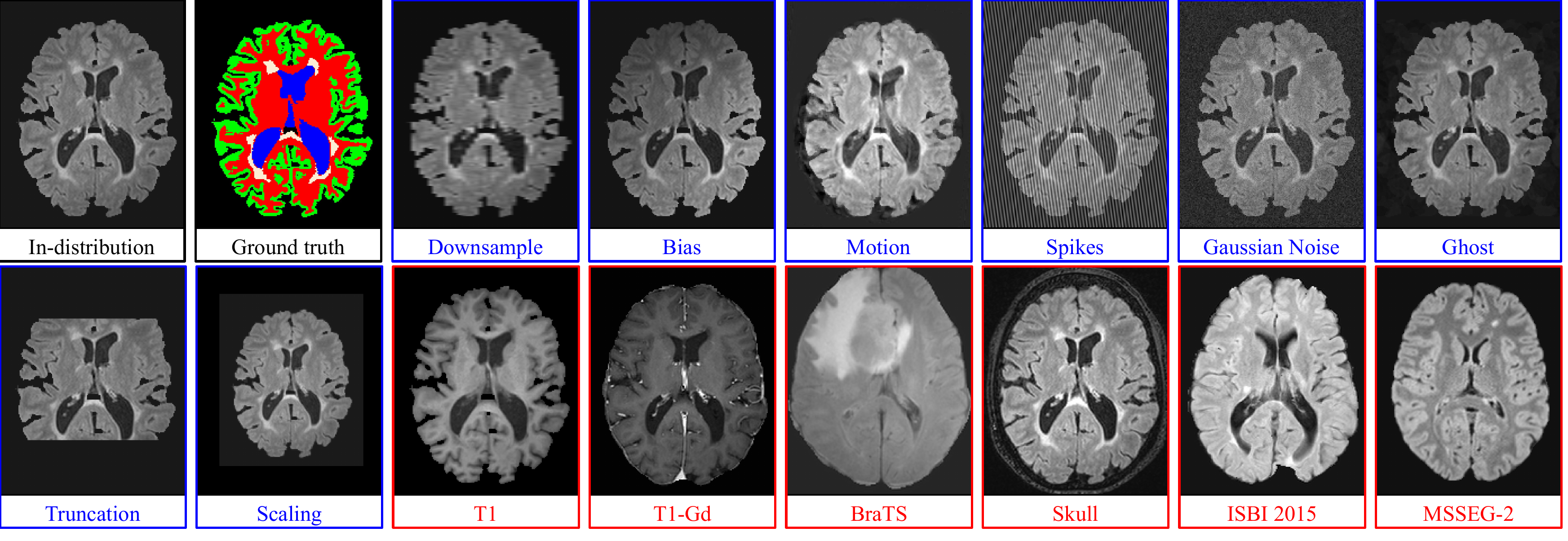}
\caption{In and Out-of-distribution data used in the experiments. In black: ID data and associated ground truth; blue: synthetic OOD; red: real OOD.}\label{fig:ood}
\end{figure*}

\section{Materials}

Data were collected to train the segmentation models, to evaluate the OOD detection, as well as to evaluate the segmentation performance of the models. 

\subsection{In-distribution data}\label{ID_Data}
We collect data from several open-source MS datasets: the MSSEG-1 Challenge \cite{commowick2016msseg}, the WMH 2017 Challenge \cite{kuijf2019standardized}, and the database gathered by the University of Ljubljana, Slovenia (MSLUB) \cite{lesjak2018novel}. For each of these datasets, T1-weighted and T2-weighted FLAIR MRI sequences are provided, as well as the ground truth segmentation of White Matter Hyperintensities (WMH).
From the total of 143 subjects, we randomly set aside 50 subjects. The remaining 93 subjects are used to train the segmentation models while the 50 set-aside subjects are included in the OOD-detection evaluation database. These 50 subjects sharing the same origin than the 93 training images, they are considered as ID images. 


We first co-register FLAIR and T1 sequences, followed by a rigid registration of both sequences on a 170 × 204 × 170 template with an isotropic resolution of 1mm. We use FastSurfer on the T1 sequences in order to obtain an anatomical atlas of the brain composed of 95 classes for each patient \cite{henschel2020fastsurfer}, which we further simplify into 6 brain regions: supratentorial white and grey matter, cerebrospinal fluid, infratentorial white and grey matter, and brain stem. We skull-strip the MRIs using the computed brain mask. We finally superpose the ground truth WMH delineations to the anatomical labels to obtain the final training masks composed of 7 foreground classes (6 anatomical classes plus 1 lesion class). Only the FLAIR sequences are used for the training of the segmentation models. 

\subsection{Out-of-distribution data}\label{ood_data}
OOD data were collected or generated to build the OOD detection evaluation database (Figure \ref{fig:ood}). The same data were also used to evaluate the segmentation performance of the models, together with the 50 set-aside ID samples. 

\subsubsection{Synthetic Out-of-distribution data}\label{syn_ood}
A common type of OOD in MRI consists is artifacts. However, finding real images with a controlled amount of artifacts to allow evaluation of OOD detection methods is difficult. We therefore generate realistic synthetic artifacted images from the set of ID evaluation images. We used the TorchIO Data Augmentation library \cite{perez2021torchio} to generate \emph{Downsample}, \emph{Bias}, \emph{Motion}, \emph{Spikes}, \emph{Gaussian Noise}, \emph{Ghost}, and \emph{Scaling} artifacts. We also implement a \emph{Truncation} operation that randomly removes the top and bottom slices of the MRI to mimic errors in the file transfer. From each ID evaluation image, we generate one OOD image per type of synthetic artefact. Having implemented 8 types of synthetic artifacts, we obtain 400 synthetically artifacted images. 

\subsubsection{Real Out-of-distribution data}\label{real_ood}
We also propose several real-world OOD scenarios to validate our approaches. We use the set-aside T1-weighted sequences of the 50 test subjects, as well post-contrast T1-weighted (T1-Gd) sequences from the MSSEG-1 dataset to simulate mistakes in the input MRI sequences of an automated pipeline (T1 or T1-Gd sequences used as input, instead of expected FLAIR series). We use 50 FLAIR sequences from the BraTS 2020 dataset \cite{menze2014multimodal} containing brain tumors to test how the models behave on a pathology not observed during training. We also construct a version of the test dataset for which no skull-stripping is applied to mimic an error in the image preprocessing pipeline. Lastly, we use two other MS datasets that are not used during training to simulate domain-shift due to different acquisition protocols. We used 21 scans from the ISBI 2015 MS dataset \cite{carass2017longitudinal} and 60 scans from the MSSEG-2 dataset \cite{commowick2021msseg}. All data follow the same preprocessing steps presented in \ref{ID_Data}, except for the samples which are deliberately not skull-stripped.

\section{Method}
We train standard and MC dropout segmentation models in 2 settings: binary (with only WMH being segmented) and multiclass (WMH as well as the 6 brain anatomical regions are segmented). MSP is obtained from the softmax probabilities of the standard models, and DUM is implemented using the features of their penultimate layers \cite{karimi2020improving}. Each type of segmentation model is trained 5 times, allowing to obtain robust statistics for MSP, MC Dropout and DUM models, as well as to build DE models. 

While DUM directly provides image-wise scores, the other implemented UQ methods (MSP, MC dropout and DE) provide one uncertainty score per voxel. For these methods, we compute the average of all the volume voxel uncertainties (estimated using the variance for MC dropout and DE) to get a single score per MRI \cite{karimi2020improving}. 

We define OOD detection as a binary classification task, with ID images defined as negative samples, and OOD images as positive samples \cite{hendrycks2016baseline}. For each OOD dataset, we compute the image-level uncertainty scores and compare them with the scores obtained for the ID set of test images. We use the area under the ROC curve (AUROC) to estimate the ability to distinguish between ID and OOD images based on the image-level uncertainty scores. When the ground truth of WMH is available (i.e. not for the BraTS and MSSEG-2 data), we also report the WMH segmentation performances of MSP, MC dropout and DE, evaluated using the Dice score (DSC).

\begin{table*}[t!]
\resizebox{\textwidth}{!}{
\begin{tabular}{|l|cc|cc|cc|cc|cc|cc|c|c|}
\hline
& \multicolumn{4}{c|}{MSP} & \multicolumn{4}{c|}{MC dropout} & 	\multicolumn{4}{c|}{DE} & \multicolumn{2}{c|}{DUM} \\

 & \multicolumn{2}{c|}{Binary} & \multicolumn{2}{c|}{Multiclass} & \multicolumn{2}{c|}{Binary} & \multicolumn{2}{c|}{Multiclass} & \multicolumn{2}{c|}{Binary} & \multicolumn{2}{c|}{Multiclass} & Binary & Multiclass \\
  & AUROC  & DSC & AUROC  & DSC & AUROC & DSC & AUROC & DSC & AUROC & DSC & AUROC & DSC & AUROC & AUROC \\ \cline{2-15}
  
Test  & -  & 0.67 $\pm$ 0.00  & -   & 0.68 $\pm$ 0.00 & - & 0.66 $\pm$ 0.01 & - & 0.66 $\pm$ 0.01 & - & 0.68   & -  & 0.68 & -   & - \\ 

Downsample  & 0.50 $\pm$ 0.02  & 0.62 $\pm$ 0.00  & \textbf{0.94 $\pm$ 0.01} & 0.63 $\pm$ 0.00 & 0.52 $\pm$ 0.02 & 0.61 $\pm$ 0.01 & 0.84 $\pm$ 0.03  & 0.62 $\pm$ 0.01 & 0.53 & 0.63 & 0.88  & 0.64 & 0.88 $\pm$ 0.03 & \textbf{0.94 $\pm$ 0.02} \\ 

Bias  & 0.58 $\pm$ 0.11  & 0.55 $\pm$ 0.01  & 0.74 $\pm$ 0.02 & 0.59 $\pm$ 0.01 & 0.51 $\pm$ 0.04  & 0.50 $\pm$ 0.02 & 0.71 $\pm$ 0.02  & 0.58 $\pm$ 0.02 & 0.58 & 0.56 & 0.74 & 0.59 & \textbf{0.94 $\pm$ 0.01} & 0.92 $\pm$ 0.02 \\ 

Motion  & 0.29 $\pm$ 0.16  & 0.63 $\pm$ 0.00 & \textbf{0.99 $\pm$ 0.01} & 0.64 $\pm$ 0.01 & 0.40 $\pm$ 0.01  & 0.63 $\pm$ 0.01 & 0.91 $\pm$ 0.03 & 0.63 $\pm$ 0.01 & 0.46 & 0.64 & 0.96 & 0.65 & 0.98 $\pm$ 0.02 & \textbf{0.99 $\pm$ 0.00} \\ 

Spikes & 0.15 $\pm$ 0.16 & 0.19 $\pm$ 0.04  & \textbf{1.00 $\pm$ 0.00} & 0.15 $\pm$ 0.04 & 0.51 $\pm$ 0.14  & 0.31 $\pm$ 0.05 & \textbf{1.00 $\pm$ 0.00}  & 0.26 $\pm$ 0.07 & 0.57 & 0.18   & \textbf{1.00} & 0.18 & \textbf{1.00 $\pm$ 0.00} & \textbf{1.00 $\pm$ 0.00} \\ 

Noise & 0.40 $\pm$ 0.36  & 0.64 $\pm$ 0.00  & 0.98 $\pm$ 0.01    & 0.64 $\pm$ 0.01  & 0.38 $\pm$ 0.03  & 0.64 $\pm$ 0.01 & 0.93 $\pm$ 0.05  & 0.64 $\pm$ 0.02 & 0.57 & 0.65 & 0.97 & 0.65  & \textbf{1.00 $\pm$ 0.00} & 0.99 $\pm$ 0.00 \\ 

Ghost & 0.41 $\pm$ 0.20 & 0.65 $\pm$ 0.00 & 0.97 $\pm$ 0.03 & 0.65 $\pm$ 0.01 & 0.48 $\pm$ 0.02  & 0.64 $\pm$ 0.01 & 0.94 $\pm$ 0.03  & 0.64 $\pm$ 0.01 & 0.52  & 0.66 & 0.96 & 0.66 & 0.96 $\pm$ 0.01 & \textbf{0.99 $\pm$ 0.04} \\ 

Truncation & 0.56 $\pm$ 0.03  & 0.64 $\pm$ 0.00  & 0.15 $\pm$ 0.01  & 0.63 $\pm$ 0.00  & 0.52 $\pm$ 0.01  & 0.62 $\pm$ 0.01 & 0.49 $\pm$ 0.04  & 0.62 $\pm$ 0.01 & 0.52 & 0.65 & 0.67 & 0.65 & 0.89 $\pm$ 0.01 & \textbf{0.96 $\pm$ 0.01} \\ 

Scale & 0.48 $\pm$ 0.11  & 0.61 $\pm$ 0.00  & 0.95 $\pm$ 0.03 & 0.61 $\pm$ 0.01 & 0.59 $\pm$ 0.03  & 0.59 $\pm$ 0.01 & 0.95 $\pm$ 0.01  & 0.60 $\pm$ 0.01 & 0.58    & 0.62   & \textbf{0.99} & 0.63 & \textbf{0.99 $\pm$ 0.00} & \textbf{0.99 $\pm$ 0.00} \\ 

T1 & 0.46 $\pm$ 0.21  & 0.00 $\pm$ 0.00  & \textbf{1.00} $\pm$ 0.00  & 0.00 $\pm$ 0.00    & 0.78 $\pm$ 0.12  & 0.00 $\pm$ 0.00 & 0.99 $\pm$ 0.00  & 0.00 $\pm$ 0.00 & 0.86    & 0.00   & \textbf{1.00} & 0.00 & 0.96 $\pm$ 0.0  & 0.99 $\pm$ 0.00            \\ 

T1-Gd & 0.31 $\pm$ 0.20  & 0.00 $\pm$ 0.00  & \textbf{1.00} $\pm$ 0.00 & 0.00 $\pm$ 0.00 & 0.42 $\pm$ 0.12 & 0.00 $\pm$ 0.00 & 1.00 $\pm$ 0.00  & 0.00 $\pm$ 0.00 & 0.49 & 0.00 & \textbf{1.00} & 0.00 & 0.98 $\pm$ 0.01  & \textbf{1.00 $\pm$ 0.00} \\ 

BraTS & 0.54 $\pm$ 0.17  & - & \textbf{1.00} $\pm$ 0.00 & - & 0.93 $\pm$ 0.02  & - & \textbf{1.00 $\pm$ 0.00}  & - & 0.96   & - & \textbf{1.00} & - & 0.99 $\pm$ 0.00 & \textbf{1.00 $\pm$ 0.00} \\ 

Skull  & 0.35 $\pm$ 0.17 & 0.51 $\pm$ 0.01  & \textbf{1.00 $\pm$ 0.00} & 0.52 $\pm$ 0.02 & 0.58 $\pm$ 0.06 & 0.49 $\pm$ 0.01 & \textbf{1.00 $\pm$ 0.00} & 0.54 $\pm$ 0.02 & 0.68 & 0.53 & \textbf{1.00} & 0.56 & 0.97 $\pm$ 0.01 & 0.99 $\pm$ 0.00 \\ 

ISBI 2015 & 0.34 $\pm$ 0.20 & 0.62 $\pm$ 0.01 & \textbf{1.00 $\pm$ 0.00} & 0.65 $\pm$ 0.02 & 0.46 $\pm$ 0.05 & 0.65 $\pm$ 0.02 & 0.99 $\pm$ 0.01 & 0.65 $\pm$ 0.02 & 0.45 & 0.62 & \textbf{1.00}  & 0.67 & 0.97 $\pm$ 0.02 & 0.99 $\pm$ 0.00  \\ 

MSSEG-2 & 0.62$\pm$ 0.05  & - & 0.94 $\pm$ 0.01 & - & 0.51 $\pm$ 0.02 & - & 0.93 $\pm$ 0.04 & - & 0.50 & - & \textbf{0.99} & - & 0.95 $\pm$ 0.02 & \textbf{0.99 $\pm$ 0.00} \\ \hline

\end{tabular}}
\caption{AUROC and DSC scores obtained for each ID and OOD datasets. For all methods except DE, we present the performance averaged over the 5 runs, as well as the standard deviation. The best OOD detection performance for each dataset is highlighted in \textbf{bold}. }\label{tab:results}
\end{table*}

\section{Implementation details}
Our framework is implemented using PyTorch \cite{paszke2019pytorch}. We use a 3D Residual UNet \cite{kerfoot2018left} as the backbone of the segmentation models. For MC dropout, we apply a 3D dropout with a rate of $20\%$ in each layer of the encoder and decoder. The models are trained using 3D patches of $160\times192\times32$ with the Dice loss \cite{milletari2016v}. We use a simple Data Augmentation pipeline during training composed of flipping, rotation, and contrast alteration. At inference, we perform $T=20$ inferences per image to obtain the final segmentation and uncertainty estimates for MC dropout models. 

\section{Results}
The AUROC and DSC scores for each OOD dataset and UQ method are presented in Table \ref{tab:results}. As expected, DSC scores are inferior on OOD data (\emph{e.g.} \emph{T1-Gd} dataset), meaning that the inability to detect OOD could result in catastrophic outcomes in the context of a real world automated pipeline. This is the case with binary implementations of MSP, MC dropout and DE that reach low OOD detection performance on most datasets, being even worst than chance ($\text{AUROC} < 0.5$) in some settings (\emph{e.g.}, MSP Binary on the \emph{Spikes} dataset). Exceptions are made for the MC dropout Binary and DE Binary that show satisfying OOD detection on the T1 and BraTS datasets, which are however the most obvious types of OOD. On the contrary, the multiclass versions demonstrate high improvement with respect to the OOD task, being able to perfectly distinguish between ID and OOD samples in many settings (\emph{e.g.}, $\text{AUROC}=1.00$ for DE Multiclass on the \emph{Skull} dataset). Interestingly, DUM models reach high detection accuracy in both settings, binary and multiclass. 


\section{Discussion and Conclusion}
In this work, we investigate the use of UQ approaches to detect OOD inputs in the context of medical image segmentation. We elaborate an evaluation framework with a set of 14 various synthetic or real OODs to evaluate several popular UQ methods for OOD detection. Our experimental results demonstrate that standard UQ approaches such as MSP, MC dropout and DE fail to detect OOD inputs when applied on binary segmentation models, even on extreme OOD cases (\emph{e.g.} \emph{T1-Gd} dataset). Overall, binary segmentation models tend to be overconfident on abnormal inputs. 

Second, our experiments show that UQ methods applied on models that were trained to segment anatomical labels additionally to lesions (multiclass) achieve excellent OOD detection. Intuitively, this additional segmentation task enforces the model to take into account the entirety of the available brain information to make its prediction, increasing the quantity of information available to decide whether the input is abnormal or not. 

Finally, DUM models show excellent OOD detection performances, with DUM Multiclass being a little more consistent than the Binary version. This suggests that the hidden activations of a trained model are a powerful indicator of the distance of a sample to the training set. Interestingly, DUM is very computationally efficient, as they only require a single trained model, in contrast to the MC dropout and DE that require either performing multiple predictions with the same network, or training several models. These approaches are thus particularly interesting for the implementation of OOD-detection tasks in a clinical setting.

While multiclass seems beneficial for OOD detection, it requires the generation of anatomical atlases to train the model, which may not always be feasible. In that case, the DUM approach is the most suitable, as it achieves strong OOD detection results even on binary models.

Future work will investigate the impact of the number of learned anatomical labels on the OOD detection and compare the performance of uncertainty-based method to the performance of dedicated models such as AE or GANs. We will also investigate the impact of preprocessing choices (\emph{e.g.} skullstripping, debiasing) on OOD detection performance.

\section{Acknowledgments}
BL, AT, SD are employees of the Pixyl company. MD and FF serve on Pixyl scientific advisory board.

\section{Compliance with Ethical Standards}
This research study was conducted retrospectively using human subject data made available in open access. Ethical approval was not required as confirmed by the license attached with the open access data.

\bibliographystyle{IEEEbib}
\bibliography{strings,refs}

\end{document}